\pdfoutput=1
\documentclass[11pt,twoside,a4paper,cmspaper,final,collab]{cms-tdr}
\def\svnVersion{1561ccf}\def\svnDate{2026/05/15}\def\cmsCernNoTag{CERN-EP-2025-280}\def\cmsCernDate{\today}\def\cmsMessage{Published in Physics Letters B as \href{http://dx.doi.org/10.1016/j.physletb.2026.140534}{\doi{10.1016/j.physletb.2026.140534}.}}
\begin{document}\cmsNoteHeader{HIG-24-017}

\cmsNoteHeader{HIG-24-017}

\newcommand{\msd}{\ensuremath{m_\mathrm{SD}}\xspace}
\newcommand{\fxfx}{{\textsc{FxFx}}\xspace}
\DeclareRobustCommand{\PX}{{\HepParticle{X}{}{}}\Xspace}

\newcommand{\Dbvqcd}{\ensuremath{\mathcal{D}(\bbbar\text{ vs. QCD})}\xspace}
\newcommand{\Dbvcq}{\ensuremath{\mathcal{D}(\bbbar\text{ vs. }\ccbar,\qqbar)}\xspace}
\newcommand{\Vhad}{\ensuremath{\PV(\Pq\Pq)}\xspace}
\newlength\cmsTabSkip\setlength{\cmsTabSkip}{1ex}

\title{Search for a boosted Higgs boson decaying to bottom quark pairs in association with a \texorpdfstring{\PW or \PZ}{W or Z} boson in proton-proton collisions at \texorpdfstring{$\sqrt{s} = 13\TeV$}{sqrt s = 13 TeV}}

\date{\today}

\abstract{
A search is conducted for standard model Higgs bosons with large transverse momentum (\pt) decaying to bottom quark pairs and produced in association with a hadronically decaying \PW or \PZ boson at the LHC.
The result is based on a dataset of proton-proton collisions at a center-of-mass energy of 13\TeV collected with the CMS detector in 2016--2018, corresponding to an integrated luminosity of 138\fbinv. Boosted Higgs, \PW, and \PZ boson decays are reconstructed using large-radius jets with $\pt>450\GeV$ and identified with heavy-flavor classifiers based on a graph convolutional neural network. The observed signal strength relative to the standard model expectation is $\mu=0.72^{+0.75}_{-0.71}$ including statistical and systematic uncertainties.
}

\hypersetup{
pdfauthor={CMS Collaboration},
pdftitle={Search for a boosted Higgs boson decaying to bottom quark pairs in association with a W or Z boson in proton-proton collisions at \texorpdfstring{$\sqrt{s} = 13\TeV$}{sqrt s = 13 TeV}},
pdfsubject={CMS},
pdfkeywords={CMS, Higgs boson, vector boson fusion}
}

\maketitle

\section{Introduction}
\label{sec:intro}

The observation of a Higgs boson (\PH) consistent with the standard model (SM) expectation in 2012~\cite{ATLAS:2012yve,CMS:2012qbp,CMS:2013btf} and subsequent measurements of its properties~\cite{CMS:2022dwd, ATLAS:2022vkf} have advanced our understanding of electroweak (EW) symmetry breaking and the origin of particle masses~\cite{Salam:1968rm,Glashow:1961tr,Weinberg:1967tq,PhysRevLett.13.321,Higgs:1964ia,PhysRevLett.13.508,PhysRev.145.1156,PhysRevLett.13.585}.
The \PH has been observed at the CERN LHC in its main production processes and several decay channels.

The dominant \PH production mechanism at the LHC is gluon fusion (ggH), contributing about 87\% of the inclusive \PH cross section at ${\sqrt{s}=13\TeV}$. However, its relative contribution decreases with increasing \pt~\cite{Becker:2020rjp}.
In contrast, the relative fraction of vector boson fusion (VBF) shows minimal \pt dependence, while the relative fraction of associated production with vector bosons ($\PV\PH$, ${\PV=\PW}$~or~\PZ) increases at high \pt.
Therefore, precise measurements of the $\PV\PH$ process could uncover beyond-the-SM (BSM) phenomena in the \PH sector, which may appear more prominently at high \pt~\cite{Brehmer:2019gmn, Biekotter:2014gup, Mimasu:2015nqa}.

The decay of an \PH into a bottom quark-antiquark pair ($\PH\to\bbbar$) is particularly well suited for studying boosted (high-\pt) \PH production because of its large branching fraction of 58\%~\cite{LHCHiggsCrossSectionWorkingGroup:2013rie}. 
Furthermore, at high \pt, the $\PH\to\bbbar$ decay products are reconstructed as a single large-radius jet, enabling the use of jet substructure and flavor-tagging observables to identify the boosted \PH candidate and suppress the dominant multijet background.

Searches for $\PV\PH$ production with $\PH\to\bbbar$ have been performed at the LHC in leptonic \PV decay channels~\cite{CMS:2018nsn,ATLASVH}, while the hadronic \PV decay channel has so far been explored only by ATLAS~\cite{ATLAS:2023jdk}; this analysis constitutes the first such search at CMS. Furthermore, this channel has been used to constrain BSM phenomena within an effective field theory framework \cite{CMS:2023vzh}. Other boosted $\PH\to\bbbar$ measurements have targeted the ggH~\cite{ATLAS:2021tbi,CMSinclusiveHbb} and VBF~\cite{CMS:2024ddc} processes. The excess of 2.7 standard deviations ($\sigma$) with respect to SM predictions observed in the previous VBF analysis~\cite{CMS:2024ddc} motivates further exploration of the $\PV\PH$ process, as it could be explained by enhanced couplings of the \PH to vector bosons.

This work extends CMS high-\pt \PH measurements in the $\PH\to\bbbar$ channel to $\PV\PH$ production in final states with a hadronic \PW or \PZ decay, \Vhad. The search uses a dataset of proton-proton (pp) collisions at $\sqrt{s}=13\TeV$ collected by the CMS experiment at the LHC in 2016--2018, corresponding to an integrated luminosity of 138\fbinv~\cite{cmsLumi16,CMS-PAS-LUM-17-004,CMS-PAS-LUM-18-002}. The same strategy is also used to measure $\Vhad\PZ(\bbbar)$ production simultaneously as a validation of the method.

\section{The CMS detector}

The CMS apparatus~\cite{CMS:2008xjf,CMS:2023gfb} is a multipurpose, nearly hermetic detector, designed to trigger on~\cite{CMS:2020cmk,CMS:2016ngn,CMS:2024aqx} and identify electrons, muons, photons, and hadrons~\cite{CMS:2020uim,CMS:2018rym,CMS:2014pgm}. Its central feature is a superconducting solenoid of 6\unit{m} internal diameter, providing a magnetic field of 3.8\unit{T}. Within the solenoid volume are a silicon pixel and strip tracker, a lead tungstate crystal electromagnetic calorimeter, and a brass and scintillator hadron calorimeter, each composed of a barrel and two endcap sections. Forward calorimeters extend the pseudorapidity ($\eta$) coverage provided by the barrel and endcap detectors. Muons are reconstructed using gas-ionization detectors embedded in the steel flux-return yoke outside the solenoid. More detailed descriptions of the CMS detector, together with a definition of the coordinate system used and the relevant kinematic variables, can be found in Refs.~\cite{CMS:2008xjf,CMS:2023gfb}.

A two-tiered trigger system selects events of interest. The first level, implemented in hardware, reduces the rate to about 100\unit{kHz} within 4\mus~\cite{CMS:2020cmk}. The high-level trigger, implemented in software, further reduces the rate to about 1\unit{kHz} before data storage~\cite{CMS:2016ngn,CMS:2024aqx}.

The dataset used here is divided into four periods: early 2016, late 2016, 2017, and 2018. These divisions reflect the 2017 tracking detector upgrade, variations in accelerator and detector conditions, and updates to reconstruction and calibration procedures.
The 2016 data are further split into early and late periods to account for a mid-year change in strip-tracker operating conditions that significantly affected tracking performance \cite{CMS-DP-2020-045}. 

\section{Simulated samples}
\label{sec:samples}
Signal and background events are simulated using various Monte Carlo (MC) event generators, with the CMS detector response modeled by \GEANTfour~\cite{GEANT4}. Independent samples are produced for each data-taking period using identical generator settings and incorporating changes in running conditions.

The background from quantum chromodynamics (QCD) multijet processes, \ie events consisting of multiple hadronic jets produced via the strong interaction, is generated at leading order (LO) accuracy with \MGvATNLO~2.6~\cite{Alwall:2014hca, Frixione:2002ik} using up to four partons in the matrix element calculation. The $\PW$+jets and $\PZ$+jets processes are also simulated at LO with \MGvATNLO, including decays to all flavors of quarks and up to three and four additional partons at the matrix element level for $\PW$+jets and $\PZ$+jets, respectively. The samples are corrected as a function of the boson \pt to match higher-order predictions. The QCD next-to-LO (NLO) corrections are derived with \MGvATNLO, simulating \PW and \PZ boson production with up to two additional partons. The EW NLO corrections are taken from theoretical calculations described in Refs.~\cite{Kallweit:2014xda, Kallweit:2015dum, Kallweit:2015fta,Lindert:2017olm}.

Diboson processes ($\PW\PW$, $\PW\PZ$, and $\PZ\PZ$) are generated at LO accuracy with \PYTHIA~8.240~\cite{Sjostrand:2014zea} and corrected to next-to-NLO (NNLO) cross sections using \MCFM~7.0~\cite{Campbell:2010ff}. Top quark pair and single-top processes are modeled at NLO in QCD with \POWHEG~2.0~\cite{Frixione:2007vw,Alioli:2010xd,Frixione:2007nw,Frederix:2012dh,Re:2010bp}.

The ggH process is simulated with HJMINLO~\cite{Monni:2019whf,Monni:2020nks}, including finite top quark mass effects~\cite{Neumann:2018bsx} as recommended in Ref.~\cite{Becker:2020rjp}. The \POWHEG~v2 generator~\cite{Nason:2009ai,Luisoni:2013kna,Hartanto:2015uka} is used for VBF \PH production, $\PV\PH$, and $\ttbar\PH$ at NLO. The VBF sample is reweighted to NNLO in the \pt spectrum and next-to-NNLO accuracy in the inclusive cross section~\cite{Cacciari:2015jma,Dreyer:2016oyx}. These corrections have a negligible effect on the yield of this process for events with \PH ${\pt>450\GeV}$. Electroweak corrections are applied to the VBF, $\PV\PH$, and $\ttbar\PH$ samples~\cite{Becker:2020rjp}.

Parton showering and hadronization are modeled with \PYTHIA using the CP5 tune~\cite{CMS:2019csb}. Jets from the parton shower are matched to those from the matrix element calculations using the MLM~\cite{Alwall:2007fs} and \fxfx~\cite{Frederix:2012ps} schemes for LO and NLO samples, respectively.
The NNPDF3.1~\cite{NNPDF:2017mvq} set of parton distribution functions (PDF) at NNLO accuracy is used for all processes. To account for additional $\Pp\Pp$ interactions occurring in the same or nearby LHC bunch crossings (pileup), the \PYTHIA generator is employed. All MC samples include effects of pileup by overlaying simulated minimum bias events on the hard-scattering event, with the multiplicity distribution matching that observed in data.

\section{Event reconstruction and selection}
\label{sec:event_reconstruction_selection}

A particle-flow (PF) algorithm~\cite{CMS:2017yfk} aims to reconstruct and identify each individual particle in an event (PF candidate) using an optimized combination of information from the various elements of the CMS detector. The primary vertex is defined as the one most consistent with the hardest scattering in the event, as determined from tracking information~\cite{CMS:2014pgm, CMS-TDR-15-02}.

Jets are formed from PF candidates using the anti-\kt clustering algorithm~\cite{Cacciari:2008gp}, as implemented in \FASTJET~\cite{Cacciari:2011ma}.
Small-radius (AK4) jets use a distance parameter of 0.4, and large-radius (AK8) jets use 0.8~\cite{CMS-PAS-JME-16-003}. The AK8 jets are used to reconstruct and select the boosted \PH and \PV boson candidates through their collimated hadronic decays, while AK4 jets enter event-level kinematic selections such as the \HT trigger requirement and jet multiplicity vetoes.
Jet momentum is calculated as the vector sum of all particle momenta in the jet.

Pileup interactions can contribute additional tracks and calorimetric energy depositions, increasing the apparent jet momentum. For AK4 jets, tracks identified to be originating from pileup vertices are discarded and an offset correction is applied to account for remaining contributions~\cite{Sirunyan:2020foa}. For AK8 jets, the pileup-per-particle identification algorithm~\cite{Sirunyan:2020foa,Bertolini:2014bba} is used to mitigate the effect of pileup at the reconstructed-particle level, making use of local shape information, event pileup properties, and tracking information. Jet energy corrections are derived from simulation studies so that the average measured energy of jets becomes identical to that of particle level jets. In-situ measurements of the momentum balance in dijet, $\PZ$+jet, photon+jet, and multijet events are used to determine any residual differences between the jet energy scale in data and simulation, and appropriate corrections are made~\cite{CMS:2016lmd}. Additional selection criteria are applied to each jet to remove jets potentially dominated by instrumental effects or reconstruction failures~\cite{Sirunyan:2020foa}.

Events are selected by triggers requiring either an AK8 jet with large \pt or a minimum event \HT, defined as the scalar \pt sum of all AK4 jets with $\pt>30\GeV$ and $\abs{\eta}<3$. These triggers are about 90\% efficient for jets with $450<\pt<500\GeV$, $\abs{\eta}<2.5$, and fully efficient above 500\GeV.

Offline, exactly two AK8 jets, corresponding to the boosted \PH and \PW/\PZ bosons, are required with ${\pt>450\GeV}$ and $\abs{\eta}<2.5$. They must pass tight jet identification requirements~\cite{CMS-PAS-JME-16-003} and be separated from any isolated leptons or photons by $\Delta R = \sqrt{\smash[b]{\Delta\eta\,^2 + \Delta\phi\,^2}} > 0.8$, where $\Delta\eta$ and $\Delta\phi$ are the differences in pseudorapidity and azimuthal angle, respectively.
Jets must satisfy ${-6.0 < \rho = 2\ln(\msd/\pt) < -2.1}$, where \pt is the jet transverse momentum and \msd is the groomed jet mass obtained with the soft drop algorithm, which removes soft, wide-angle radiation, using parameters $z_\text{cut}=0.1$ and $\beta=0$~\cite{Larkoski:2014wba}.
The lower $\rho$ threshold removes regions where nonperturbative effects are poorly modeled, and the upper $\rho$ threshold removes jets where the constituents are not fully contained by the jet cone.
The soft drop algorithm reduces contamination from underlying events and pileup, improving the jet mass resolution to about 10\% for boosted two-prong jets and stabilizing the mass response across a wide range of jet momenta.

To suppress the background contribution from semi- and di-leptonic \ttbar decays, events are required to contain no isolated leptons. The relative isolation variable for electrons (muons) is defined as the scalar sum \pt of charged hadrons and neutral particles within a cone of radius $\Delta R=0.3$ (0.4) around the lepton, corrected for pileup and divided by the lepton \pt. For the purpose of this veto, electrons must have $\pt>10\GeV$, $\abs{\eta}<2.5$, pass loose identification criteria~\cite{CMS:2020uim}, and have relative isolation less than 0.15. Muons must have $\pt>10\GeV$, $\abs{\eta}<2.4$, pass loose identification criteria~\cite{CMS:2018rym}, and have relative isolation less than 0.25. Hadronically decaying tau leptons, \tauh, must have $\pt>20\GeV$, $\abs{\eta}<2.3$, and pass the \textsc{DeepTau} algorithm identification requirements~\cite{CMS:2022prd}. Events are also required to have $\ptmiss < 140\GeV$. The missing transverse momentum vector \ptvecmiss is computed as the negative vector sum of the transverse momenta of all the PF candidates in an event, and its magnitude is denoted as \ptmiss~\cite{CMS:2019ctu}. The \ptvecmiss is modified to account for corrections to the energy scale of the reconstructed jets in the event. In order to reduce the contribution from fully hadronic \ttbar decays, events are also vetoed if they have six or more AK4 jets with $\pt>30\GeV$ and $\abs{\eta}<5$.

The two selected AK8 jets are designated as the \PH and \PV candidates. The flavor properties of each jet are determined using
the mass-decorrelated (MD) version of the \textsc{ParticleNet} algorithm~\cite{CMS-PAS-BTV-22-001}. This version of the tagger, denoted \textsc{ParticleNet-MD}, is trained on simulated jets with uniform distributions in mass and \pt to minimize dependence of the jet selection on its mass.
The \textsc{ParticleNet-MD} classifier assigns four scores to each AK8 jet, $p(\PX\to\bbbar)$, $p(\PX\to\ccbar)$, $p(\PX\to\qqbar)$, and $p(\text{QCD})$, corresponding to four categories: three for a boson \PX decaying to bottom, charm, or light quarks, and one for generic QCD background jets not consistent with the targeted two-prong boson-decay topology.
The scores sum to unity. Discriminants formed from these scores are used to classify the AK8 jets and to suppress the QCD background.

The \PH candidate (Jet 1) is chosen to be the jet with the higher discriminant \Dbvcq, where
\ifthenelse{\boolean{cms@external}}
{
	\begin{multline*}
		\Dbvcq = \\
		\hfill \frac{p(\PX\to\bbbar)}{p(\PX\to\bbbar) +  p(\PX\to\ccbar) + p(\PX\to\qqbar)}.
	\end{multline*}
}
{
	\begin{equation*}
		\Dbvcq = \frac{p(\PX\to\bbbar)}{p(\PX\to\bbbar) +  p(\PX\to\ccbar) + p(\PX\to\qqbar)}.
	\end{equation*}
}
The other selected AK8 jet (Jet 2) is designated as the \PV candidate. Events where the \PV candidate jet has a high $p(\text{QCD})$ score are discarded, where the $p(\text{QCD})$ threshold corresponds to a signal efficiency of 25\% and a misidentification rate of 3\% on QCD jets~\cite{CMS-PAS-BTV-22-001}. Events are divided into three categories based on the \msd of the \PV candidate jet: 40--68\GeV, 68--110\GeV, and 110--201\GeV.
The middle category contains the peak corresponding to the \PV boson, while the others serve as sideband control regions (CRs) that constrain backgrounds containing only one hadronic decay of a heavy particle.

A second flavor discriminant, \Dbvqcd, is used to determine whether the \PH candidate jet originates from a heavy particle decaying to two \PQb quarks. This discriminant is defined as a ratio of \textsc{ParticleNet-MD} classifier scores:
\begin{equation*}
   \Dbvqcd = \frac{p(\PX \to \bbbar)}{p(\PX\to\bbbar) + p(\text{QCD})}.
\end{equation*}
Events that have high \Dbvqcd, above the threshold corresponding to 40\% $\PX \to \bbbar$ signal efficiency~\cite{CMS-PAS-BTV-22-001}, enter the signal region (SR). Events with lower \Dbvqcd form a CR dominated by QCD multijet processes, which is used to estimate the background contribution from QCD.

\section{Background estimation}
\label{sec:bkg}

The QCD multijet process is the dominant background in all analysis categories. Significant resonant backgrounds arise from $\PW$+jets and $\PZ$+jets processes, where the two-prong decay of a high-$\pt$ vector boson is mistagged as a \PH candidate. The \ttbar process and single top quarks produced in association with a \PW boson form nonresonant backgrounds, where the decay of a high-$\pt$ top quark is reconstructed as a large-radius jet and passes the \PH candidate selection. Additional minor backgrounds arise from other \PH production processes (ggH, VBF, and $\ttbar\PH$). The QCD multijet and top quark backgrounds are determined with data in CRs, while the other backgrounds are estimated using simulation.

\subsection{Multijet background}
\label{sec:qcdbkg}

{\tolerance=1200
The QCD multijet background in the SR is estimated from data using events failing the \Dbvqcd requirement, referred to as the QCD CR.
The shape of the \PH candidate \msd distribution of QCD background in the SR is taken from the QCD CR and multiplied by a parametric transfer factor, $TF(\msd)$, to account for shape differences due to the \Dbvqcd requirement and differences in the tagger performance between data and simulation.
The coefficients of this transfer factor are freely floating parameters determined directly from a simultaneous maximum likelihood fit to data in the signal and control regions.
\par}

The QCD multijet background yield in the SR is
\begin{align*}
   N_\text{SR}(\msd) =  R_\text{SR/CR}^{\text{MC}} \; N_\text{CR}^{\text{data}}(\msd) \;  TF(\msd),
\end{align*}
where $R_\text{SR/CR}^{\text{MC}}$ is the ratio of the QCD MC event yield in the SR and CR, $N_\text{CR}^{\text{data}}$ is the number of data events minus the number of predicted non-QCD background events in the QCD CR,  and $TF(\msd)$ is the transfer factor that accounts for shape differences between the CR and SR. The same transfer factor is applied to all three bins in the \PV candidate \msd, allowing the sideband regions to help constrain the QCD background. Independent transfer factors are defined for each data-taking period. Each transfer factor is defined as a Bernstein polynomial~\cite{CMS:2024ddc} in $\log(\msd/\GeVns{})$, and a first-order polynomial is determined to be optimal for all data-taking periods using a Fisher F-test~\cite{ftest}. The transfer factor is therefore defined as
\ifthenelse{\boolean{cms@external}}
{
	\begin{multline*}
		TF(\msd) =\\
		a (1-\log(\msd/\GeVns{})) + b \log(\msd/\GeVns{}),
	\end{multline*}
}
{
	\begin{align*}
		TF(\msd) = a (1-\log(\msd/\GeVns{})) + b \log(\msd/\GeVns{}),
	\end{align*}
}
where the values of the coefficients $a$ and $b$ are obtained in the simultaneous fit to data described in Section~\ref{sec:results}. Alternative functional forms for $TF(\msd)$ were tested and found to have a negligible impact on the final result.

\subsection{Vector boson backgrounds}

Resonant backgrounds due to $\PV$+jets and diboson processes are estimated using simulation. A $\PW$-enriched CR, distinct from the top quark CR described later, is used to calibrate jet mass scale and resolution, and to determine the efficiency of the \textsc{ParticleNet-MD} $p(\text{QCD})$ selection requirement on the \PW candidate jet.

Events in this CR are collected with single-muon triggers. Offline, events are required to have one muon with $\pt>53\GeV$, $\abs{\eta}<2.1$, relative isolation less than 0.15, $\ptmiss>40\GeV$, no additional leptons, and an AK8 jet with $\pt>450\GeV$ passing the baseline kinematic, identification, and isolation requirements described in Section \ref{sec:event_reconstruction_selection}. The event must also contain at least one AK4 jet with $\pt>50\GeV$ and $\abs{\eta}<2.5$ that is separated by $\Delta R>0.8$ from the AK8 jet and \PQb-tagged using the \textsc{DeepJet} algorithm~\cite{Bols:2020bkb,CMS-DP-2018-058}. This CR contains a high fraction of \ttbar events where one top quark decays semi-leptonically and the other to a boosted \PW boson.

Template \msd distributions are derived from simulation in two regions based on whether the AK8 jet passes or fails the $p(\text{QCD})$ requirement. The templates in each region correspond to events that contain a \PW boson and events that do not. For the former, variations of the \msd scale and resolution are calculated using linear morphing~\cite{Baak:2014fta}.

An independent fit to data in this CR, separate from the simultaneous signal extraction described in Section \ref{sec:results}, is performed to extract three corrections.
The fit is evaluated simultaneously on the \msd distributions in regions passing and failing the $p(QCD)$ selection.
In this fit, jets successfully matched to a hadronic W decay form a resonant mass peak, which constrains the jet mass scale shift and resolution, while the relative yield of W-boson jets between the pass and fail regions determines the $p(QCD)$ selection efficiency.
The corrections are derived independently for each data-taking period.
The measured $p(\text{QCD})$ efficiency ranges from 0.70 to 0.89 with relative uncertainties of 7 to 16\%, depending on the data-taking period. The jet mass scale correction is consistent with unity, with uncertainties of 200--300\MeV. The jet mass resolution scale factor ranges from 1.14 to 1.25, with uncertainties of 0.13--0.38.

\subsection{Top quark backgrounds}

Top quark processes constitute a nonresonant component of the background.
The shapes of the \msd distributions in \ttbar and single top quark processes are estimated from simulation.
The normalizations of the \ttbar and single top quark backgrounds, as well as the efficiency of the \Dbvqcd requirement on top quark jets, are constrained using a CR targeting \ttbar events, where one top quark decays via $\PQt\to\PW(\PGm\PGn)\PQb$ and the other decays hadronically. Events in this CR are required to have exactly one muon with $\pt>55\GeV$, $\abs{\eta}<2.1$, and relative isolation less than 0.25; no electrons or \tauh; at least one AK4 jet \PQb-tagged with \textsc{DeepJet}; and one AK8 jet with $\pt>450\GeV$ and $\abs{\eta}<2.5$. The \PQb-tagged AK4 jet must be separated from the AK8 jet by $\Delta R > 0.8$. This region is about 80\% pure in \ttbar, with small contributions from QCD, $\PW$+jets, $\PZ$+jets, and single top quark processes. The region is split into passing and failing \Dbvqcd categories, and both are fitted simultaneously with the SR during the signal extraction. The resulting scale factors for the normalizations and the efficiency of the \Dbvqcd requirement are consistent with unity.

\section{Systematic uncertainties}
\label{sec:systematics}

Systematic uncertainties account for possible mismodeling of the background and signal processes. They are included as nuisance parameters in the signal extraction procedure and treated according to the frequentist paradigm~\cite{CMS-NOTE-2011-005}.

The dominant systematic uncertainty arises from the modeling of the QCD multijet background, specifically from the coefficients of $TF(\msd)$ that are left unconstrained in the fit.
Additional small experimental systematic uncertainties are included, amounting to less than a few percent.
The uncertainties in the jet \msd scale and resolution derived from the \PW boson-enriched CR are propagated to all resonant processes using linear morphing of the templates, and the uncertainty in the $p(\text{QCD})$ tagging efficiency is propagated to the normalizations of the $\PV\PH$ and $\PV\PV$ processes.
The integrated luminosities for the 2016, 2017, and 2018 data-taking years have 1.2--2.5\% individual uncertainties, while the overall uncertainty is 1.6\%~\cite{cmsLumi16,CMS-PAS-LUM-17-004,CMS-PAS-LUM-18-002}. The jet energy scale and resolution uncertainties range from 0.5 to 4.0\%, and are treated as independent from the \msd uncertainties. Variations in the amount of pileup obtained by adjusting the total inelastic cross section, modeling of the trigger efficiency, and the isolation and identification of the vetoed leptons are also considered, and contribute only a few percent to the total uncertainty.

Theoretical uncertainties are included in the final fit to account for inaccuracy in the modeling of SM processes. The dominant theoretical uncertainty is due to the renormalization and factorization scales chosen for the simulated \PH and diboson samples. These uncertainties are applied to the \PH and diboson processes and are assessed by varying the scales up and down by factors of two, while keeping their ratio between 0.5 and 2.
This uncertainty is propagated to the total expected yield of the  $\PV\PH$ and $\PV\PZ$ processes according to the prescription recommended in Ref.~\cite{Becker:2020rjp}. Uncertainties in the signal processes due to initial- and final-state radiation are also calculated by varying the renormalization scale and nonsingular term using the \PYTHIA showering algorithm \cite{Mrenna:2016sih}. The Hessian PDF uncertainty is also applied to signal yield, according to the prescription of Ref.~\cite{Butterworth:2015oua}.

Theoretical uncertainties in the $\PW$+jets and $\PZ$+jets processes account for missing higher-order QCD and mixed QCD-EW effects beyond the corrections described in Section~\ref{sec:samples}, following the prescription of Ref.~\cite{Lindert:2017olm}. All theoretical uncertainties are considered to be correlated across data-taking periods, and experimental systematic uncertainties are considered to be uncorrelated, with the exception of correlated components of the uncertainties in the integrated luminosity.

Beyond the dominant uncertainty from the QCD background modeling, subleading contributions originate from jet-mass scale and resolution calibrations, as well as from the top-quark background normalization and pass/fail efficiency constrained in the muon CR.
Tagger-efficiency scale factors have smaller effects and are only weakly constrained. Theoretical and luminosity uncertainties primarily affect the overall normalization and have a negligible impact on the fitted signal strength. The precision on the signal strength is ultimately limited by the amount of data available, in particular by the finite sideband sample constraining the QCD yield beneath the \PH mass peak through the transfer factor.

\section{Results}
\label{sec:results}

A binned maximum likelihood fit to the observed \PH candidate \msd distributions is performed using the sum of the signal and background contributions. The test statistic chosen to determine the signal yield is based on the profile likelihood ratio~\cite{CMS-NOTE-2011-005}. The fit is performed simultaneously using the \Dbvqcd pass and fail regions for all three bins in the \PV boson candidate mass and the \ttbar CR, and for all four data-taking periods. The results have been obtained using the CMS statistical analysis tool \textsc{Combine}~\cite{CMS:2024onh}, which is based on the \textsc{RooFit}~\cite{Verkerke:2003ir} and \textsc{RooStats}~\cite{Moneta:2010pm} frameworks.

\begin{figure*}[htbp!]
   \centering
   \includegraphics[width=0.45\textwidth]{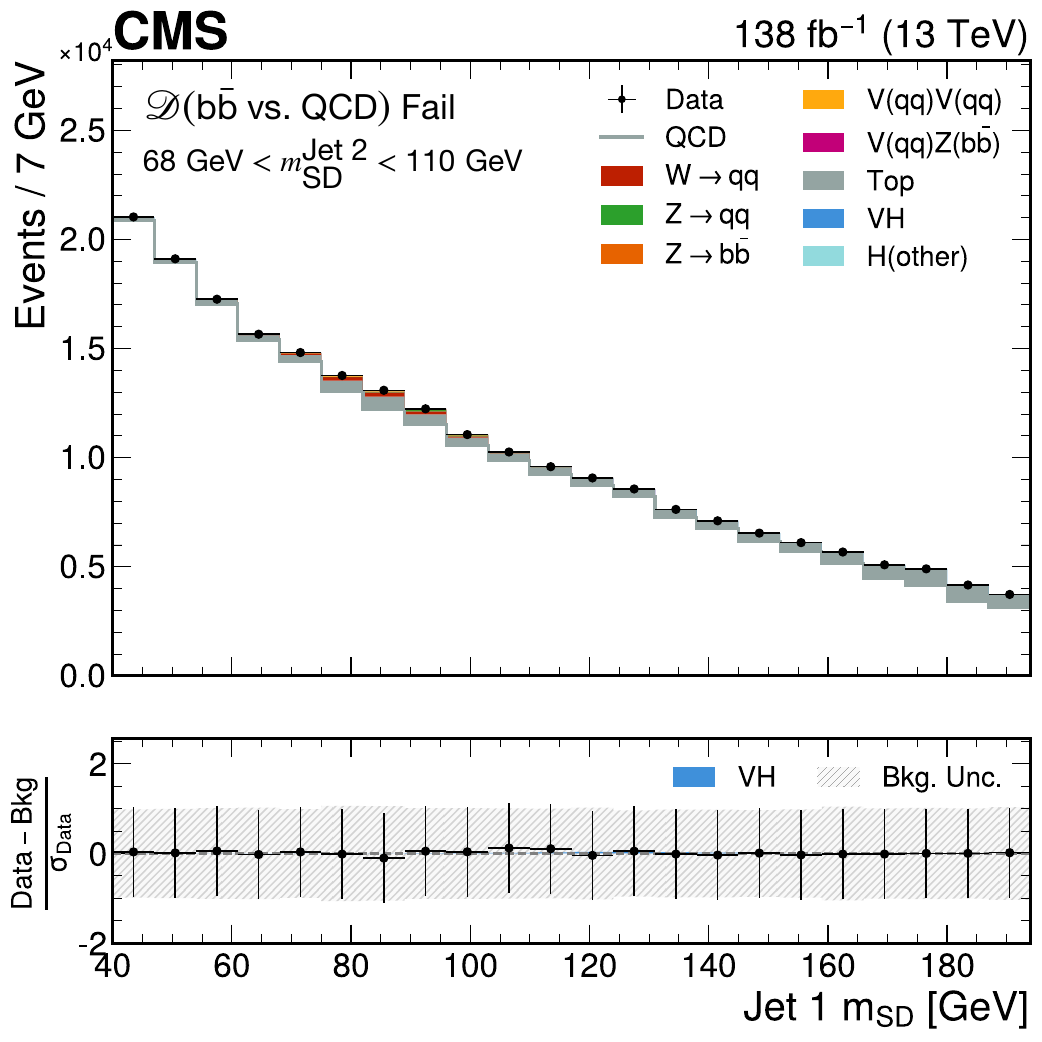}
   \includegraphics[width=0.45\textwidth]{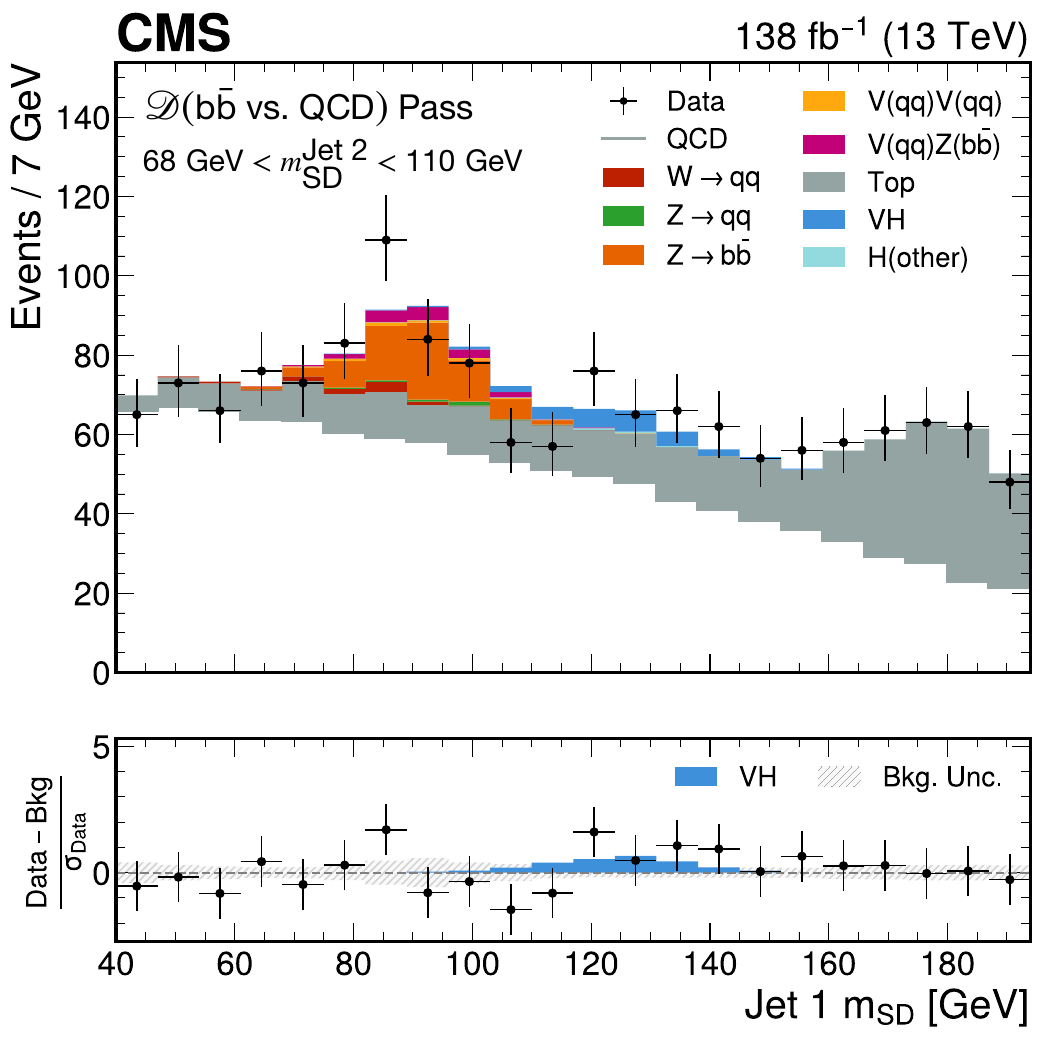}
   \caption{The \PH candidate \msd distribution in the \Dbvqcd fail (left) and pass region (right), both in the category with \PV candidate mass between 68 and 110\GeV.
   The lower panels show residuals normalized to the data statistical uncertainty, yielding unit error bars.
   Agreement in the fail region (left) is by construction, as the QCD background is derived directly from the data.
   The grey bands represent the systematic uncertainty in the background prediction. The $\Vhad\PZ(\bbbar)$ process is separated from other $\Vhad\Vhad$ processes since it is used to validate the analysis strategy.}
   \label{fig:result-vh}
\end{figure*}

\begin{figure*}[htbp!]
   \centering
   \includegraphics[width=0.45\textwidth]{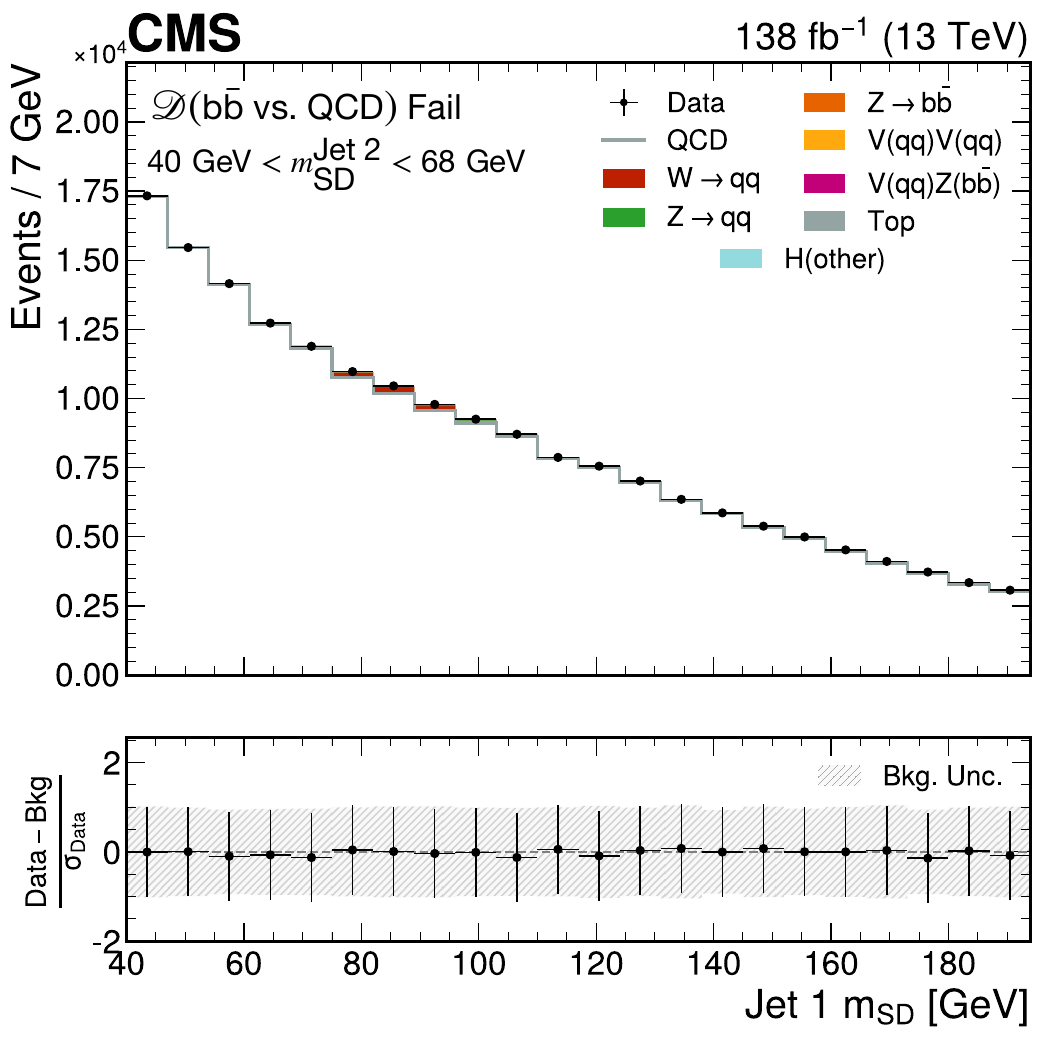}
   \includegraphics[width=0.45\textwidth]{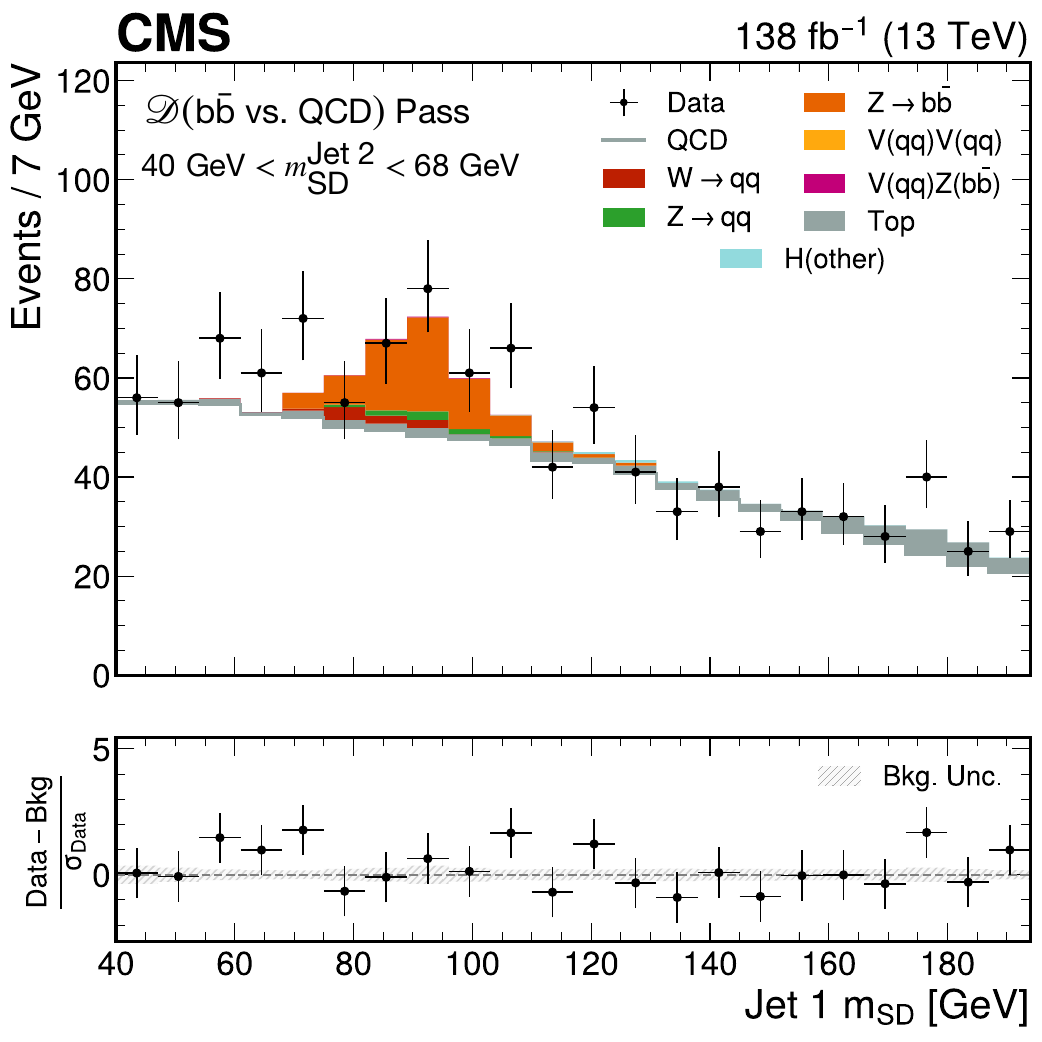}
   \\
   \includegraphics[width=0.45\textwidth]{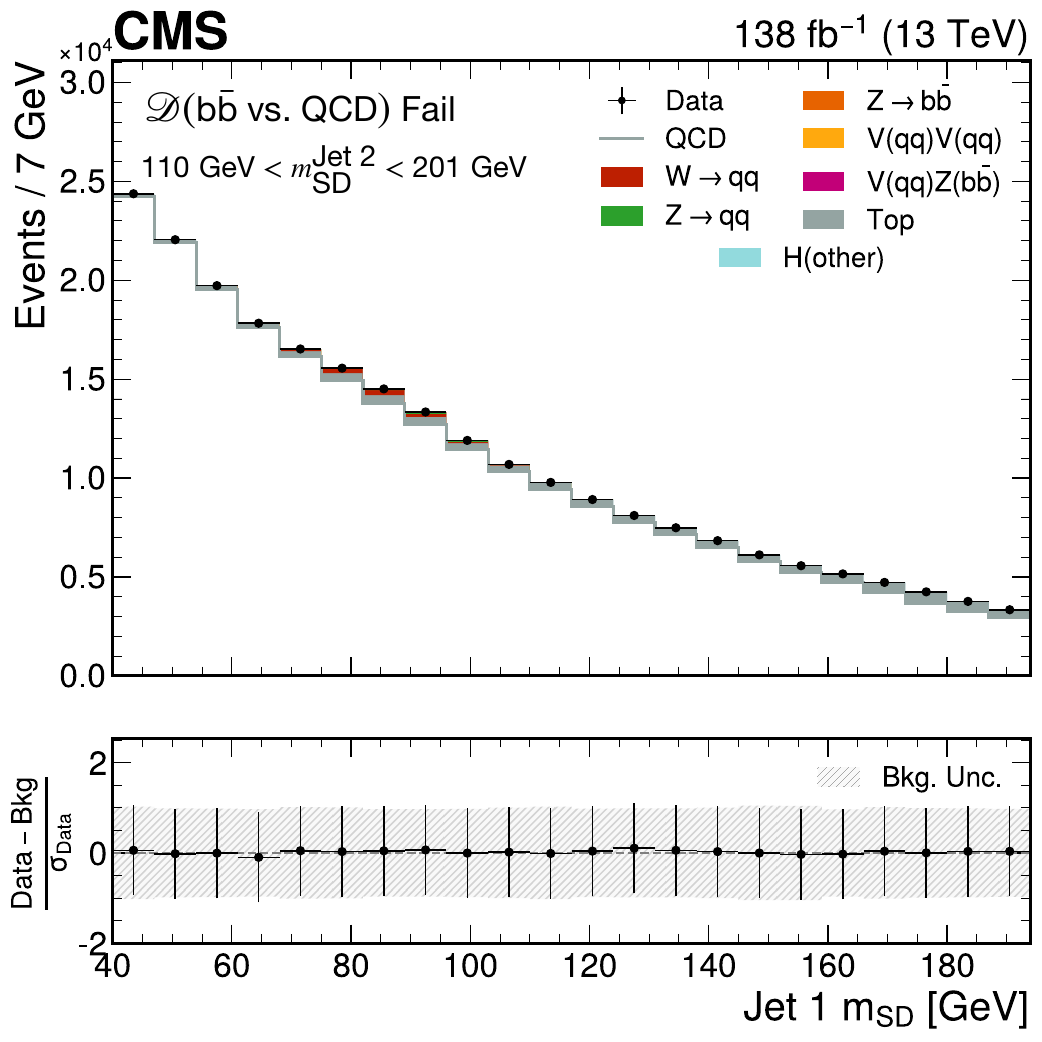}
   \includegraphics[width=0.45\textwidth]{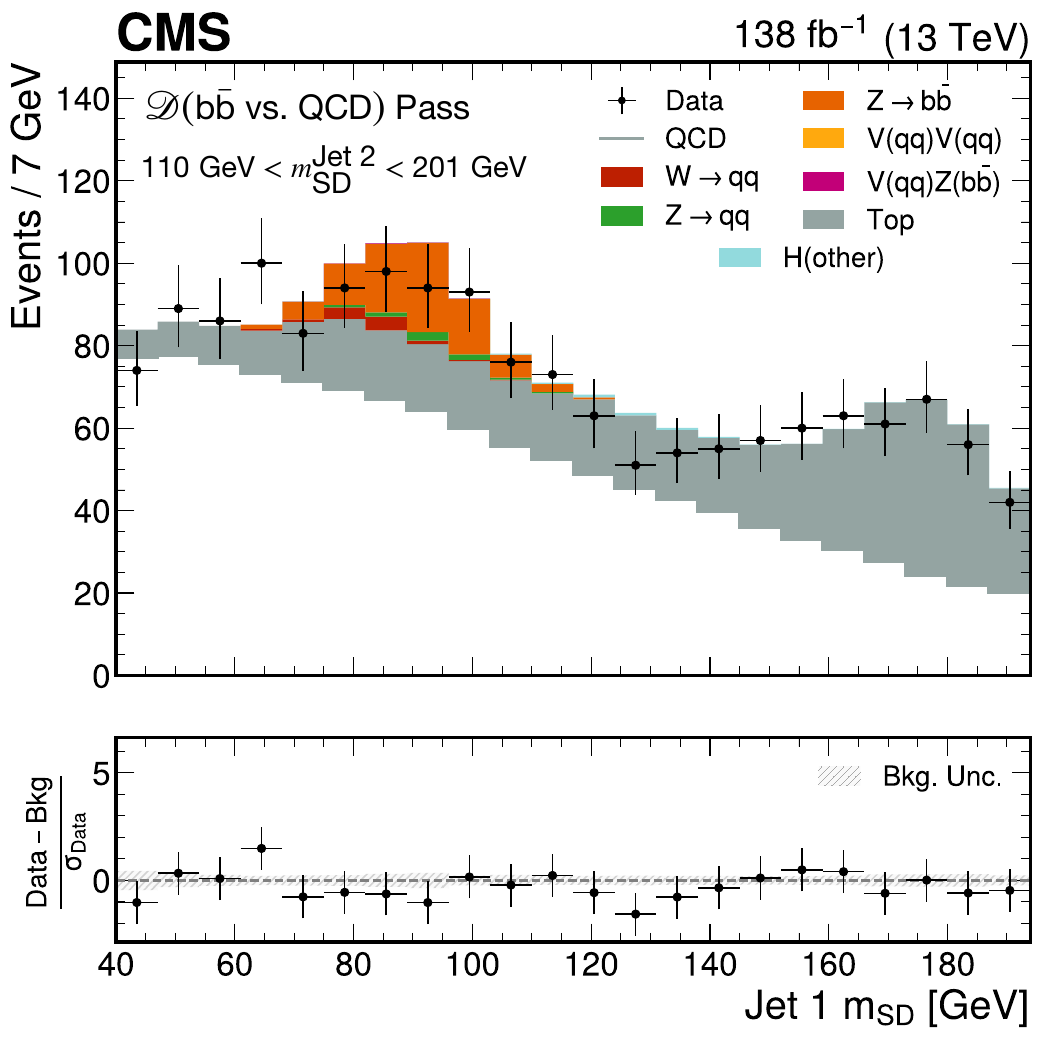}

   \caption{The \PH candidate \msd distribution in the \Dbvqcd fail (left) and pass region (right) regions, in the category where the \PV candidate \msd is in the range 40--68\GeV (upper) and 110--201\GeV (lower). Other details are as described in the caption of Fig. 1.}
   \label{fig:result-vh-sideband}
\end{figure*}

The observed data and fitted \msd distributions in the \Dbvqcd fail and pass regions are shown in Fig.~\ref{fig:result-vh} for the category where the \PV candidate has \msd between 68 and 110\GeV. The \PV mass sidebands (\PV candidate \msd in the ranges 40--68 and 110--201\GeV) are shown in Fig.~\ref{fig:result-vh-sideband}. The near-perfect model agreement with data in the fail regions is by construction: these regions dominate the QCD background determination due to the large event count compared to the pass regions.

The background in both the \Dbvqcd fail and pass regions is dominated by QCD. In the \Dbvqcd pass region, the \ttbar background exhibits a peak in the \msd of the \PH candidate near the top quark mass of 173\GeV. This peak corresponds to the decay of a highly boosted top quark, where the \PH candidate jet (Jet 1) captures most of the decay products. The large fraction of \ttbar events entering the highest \PV mass category (Fig. \ref{fig:result-vh-sideband}, lower right), where the top quark peak is especially prominent, corresponds to events where two boosted top quarks are captured in the two selected AK8 jets.

The contribution of $\PZ(\bbbar)$+jets production to the background in the \Dbvqcd pass region is clearly visible at the \PZ mass peak for all three \PV mass categories. Mistagged \PW and \PZ decays to light quarks and diboson processes also comprise a small portion of the background.
In particular, the $\Vhad\PZ(\bbbar)$ process generally satisfies the same selection criteria as the $\Vhad\PH(\bbbar)$ signal, and therefore it is used to validate the analysis strategy.

The best fit value of the $\PV\PH$ signal strength ($\mu_{\PV\PH}$), defined as the ratio of the measured to the SM expected cross section times $\PH\to\bbbar$ branching fraction, and an approximate 68\% confidence level interval are extracted following Ref.~\cite{CMS:2014fzn}. The signal strength of the $\Vhad\PZ(\bbbar)$  process ($\mu_{\PV\PZ}$) is extracted simultaneously with $\mu_{\PV\PH}$.

The fitted signal strengths for $\PV\PH$ and $\PV\PZ$ production are extracted separately for the four data-taking periods, and a combined fit over the full dataset is performed for the final result. The ggH, VBF, and $\ttbar\PH$ processes are fixed to their SM predictions. The results are \ifthenelse{\boolean{cms@external}}{}{\linebreak} summarized in Table~\ref{tab-sig}.  The observed (expected) signal strength for the $\PV\PH$ process is \ifthenelse{\boolean{cms@external}}{}{\linebreak} ${\mu_{\PV\PH}=0.72^{+0.75}_{-0.71}}$~($1.00^{+0.67}_{-0.62}$), and the observed (expected) signal strength for the $\PV\PZ$ process is ${\mu_{\PV\PZ}= 0.09^{+0.63}_{-0.63}}$~($1.00^{+0.58}_{-0.58}$).
This corresponds to an observed (expected) significance with respect to the background-only hypothesis of 1.00 (1.64) $\sigma$ for the $\Vhad\PH(\bbbar)$ process and 0.15 (1.76) $\sigma$ for the $\Vhad\PZ(\bbbar)$ process. Tabulated results are provided in the HEPData record for this analysis~\cite{hepdata}.

\begin{table}[htbp!]
\centering
\topcaption{Fitted signal strengths and their total uncertainties for the $\PV\PH$ and $\PV\PZ$ processes, shown individually for each data-taking period and for the combined dataset.}
\renewcommand{\arraystretch}{1.3}
\begin{tabular}{lcrr}
{} & $\mathcal{L}$ $[\fbinv]$ & $\mu_{\PV\PH}$ & $\mu_{\PV\PZ}$\\ \hline
Early 2016 & 19.5 & $1.02^{+2.52}_{-2.23}$  & $-0.15^{+1.77}_{-1.77}$ \\
Late 2016 & 16.8 & $1.34^{+2.20}_{-1.87}$ & $-0.83^{+1.53}_{-1.53}$ \\
2017 & 41.5 & $0.00^{+1.26}_{-1.18}$  & $1.20^{+1.28}_{-1.28}$ \\
2018 & 59.8 & $1.04^{+1.22}_{-1.10}$  & $-0.40^{+0.92}_{-0.92}$\\[\cmsTabSkip]
Combined & 138.0 & $0.72^{+0.75}_{-0.71}$ & $0.09^{+0.63}_{-0.63}$ \\
\end{tabular}
\label{tab-sig}
\end{table}

\section{Summary}

A search for standard model Higgs boson (\PH) candidates with high transverse momentum (\pt) decaying to bottom quark pairs ($\bbbar$) in association with a hadronically decaying \PW or \PZ (\PV) boson at the LHC has been presented. The analysis identifies collision events containing boosted hadronic jets with $\pt>450\GeV$ and with substructure and flavor properties characteristic of $\PH\to\bbbar$ and hadronic \PV decays. The signal strength for the $\Vhad\PH(\bbbar)$ process extracted from a maximum likelihood fit to the data in the large-radius jet mass distribution is $\mu_{\PV\PH}=0.72^{+0.75}_{-0.71}$ including statistical and systematic uncertainties. This corresponds to a significance of 1.00 standard deviations ($\sigma$), compared to 1.64 $\sigma$ expected. The sensitivity of this measurement is primarily limited by the amount of data available.

\begin{acknowledgments}
We congratulate our colleagues in the CERN accelerator departments for the excellent performance of the LHC and thank the technical and administrative staffs at CERN and at other CMS institutes for their contributions to the success of the CMS effort. In addition, we gratefully acknowledge the computing centers and personnel of the Worldwide LHC Computing Grid and other centers for delivering so effectively the computing infrastructure essential to our analyses. Finally, we acknowledge the enduring support for the construction and operation of the LHC, the CMS detector, and the supporting computing infrastructure provided by the following funding agencies: SC (Armenia), BMBWF and FWF (Austria); FNRS and FWO (Belgium); CNPq, CAPES, FAPERJ, FAPERGS, and FAPESP (Brazil); MES and BNSF (Bulgaria); CERN; CAS, MoST, and NSFC (China); MINCIENCIAS (Colombia); MSES and CSF (Croatia); RIF (Cyprus); SENESCYT (Ecuador); ERC PRG, TARISTU24-TK10 and MoER TK202 (Estonia); Academy of Finland, MEC, and HIP (Finland); CEA and CNRS/IN2P3 (France); SRNSF (Georgia); BMFTR, DFG, and HGF (Germany); GSRI (Greece); NKFIH (Hungary); DAE and DST (India); IPM (Iran); SFI (Ireland); INFN (Italy); MSIT and NRF (Republic of Korea); MES (Latvia); LMTLT (Lithuania); MOE and UM (Malaysia); BUAP, CINVESTAV, CONACYT, LNS, SEP, and UASLP-FAI (Mexico); MOS (Montenegro); MBIE (New Zealand); PAEC (Pakistan); MES, NSC, and NAWA (Poland); FCT (Portugal);MESTD (Serbia); MICIU/AEI and PCTI (Spain); MOSTR (Sri Lanka); Swiss Funding Agencies (Switzerland); MST (Taipei); MHESI (Thailand); TUBITAK and TENMAK (T\"{u}rkiye); NASU (Ukraine); STFC (United Kingdom); DOE and NSF (USA).
\hyphenation{Rachada-pisek} Individuals have received support from the Marie-Curie program and the European Research Council and Horizon 2020 Grant, contract Nos.\ 675440, 724704, 752730, 758316, 765710, 824093, 101115353, 101002207, 101001205, and COST Action CA16108 (European Union); the Leventis Foundation; the Alfred P.\ Sloan Foundation; the Alexander von Humboldt Foundation; the Science Committee, project no. 22rl-037 (Armenia); the Fonds pour la Formation \`a la Recherche dans l'Industrie et dans l'Agriculture (FRIA) and Fonds voor Wetenschappelijk Onderzoek contract No. 1228724N (Belgium); the Beijing Municipal Science \& Technology Commission, No. Z191100007219010, the Fundamental Research Funds for the Central Universities, the Ministry of Science and Technology of China under Grant No. 2023YFA1605804, the Natural Science Foundation of China under Grant No. 12535004, and USTC Research Funds of the Double First-Class Initiative No.\ YD2030002017 (China); the Ministry of Education, Youth and Sports (MEYS) of the Czech Republic; the Shota Rustaveli National Science Foundation, grant FR-22-985 (Georgia); the Deutsche Forschungsgemeinschaft (DFG), among others, under Germany's Excellence Strategy -- EXC 2121 ``Quantum Universe" -- 390833306, and under project number 400140256 - GRK2497; the Hellenic Foundation for Research and Innovation (HFRI), Project Number 2288 (Greece); the Hungarian Academy of Sciences, the New National Excellence Program - \'UNKP, the NKFIH research grants K 131991, K 133046, K 138136, K 143460, K 143477, K 146913, K 146914, K 147048, 2020-2.2.1-ED-2021-00181, TKP2021-NKTA-64, and 2025-1.1.5-NEMZ\_KI-2025-00004 (Hungary); the Council of Science and Industrial Research, India; ICSC -- National Research Center for High Performance Computing, Big Data and Quantum Computing, FAIR -- Future Artificial Intelligence Research, and CUP I53D23001070006 (Mission 4 Component 1), funded by the NextGenerationEU program (Italy); the Latvian Council of Science; the Ministry of Education and Science, project no. 2022/WK/14, and the National Science Center, contracts Opus 2021/41/B/ST2/01369, 2021/43/B/ST2/01552, 2023/49/B/ST2/03273, and the NAWA contract BPN/PPO/2021/1/00011 (Poland); the Funda\c{c}\~ao para a Ci\^encia e a Tecnologia, grant CEECIND/01334/2018 (Portugal); the National Priorities Research Program by Qatar National Research Fund;  MICIU/AEI/10.13039/501100011033, ERDF/EU, ``European Union NextGenerationEU/PRTR", and Programa Severo Ochoa del Principado de Asturias (Spain); the Chulalongkorn Academic into Its 2nd Century Project Advancement Project, the National Science, Research and Innovation Fund program IND\_FF\_68\_369\_2300\_097, and the Program Management Unit for Human Resources \& Institutional Development, Research and Innovation, grant B39G680009 (Thailand); the Kavli Foundation; the Nvidia Corporation; the SuperMicro Corporation; the Welch Foundation, contract C-1845; and the Weston Havens Foundation (USA).
\end{acknowledgments}
\bibliography{auto_generated}
\cleardoublepage \appendix\section{The CMS Collaboration \label{app:collab}}\begin{sloppypar}\hyphenpenalty=5000\widowpenalty=500\clubpenalty=5000
\cmsinstitute{Yerevan Physics Institute, Yerevan, Armenia}
{\tolerance=6000
A.~Hayrapetyan, V.~Makarenko\cmsorcid{0000-0002-8406-8605}, A.~Tumasyan\cmsAuthorMark{1}\cmsorcid{0009-0000-0684-6742}
\par}
\cmsinstitute{Institut f\"{u}r Hochenergiephysik, Vienna, Austria}
{\tolerance=6000
W.~Adam\cmsorcid{0000-0001-9099-4341}, J.W.~Andrejkovic, L.~Benato\cmsorcid{0000-0001-5135-7489}, T.~Bergauer\cmsorcid{0000-0002-5786-0293}, M.~Dragicevic\cmsorcid{0000-0003-1967-6783}, C.~Giordano, P.S.~Hussain\cmsorcid{0000-0002-4825-5278}, M.~Jeitler\cmsAuthorMark{2}\cmsorcid{0000-0002-5141-9560}, N.~Krammer\cmsorcid{0000-0002-0548-0985}, A.~Li\cmsorcid{0000-0002-4547-116X}, D.~Liko\cmsorcid{0000-0002-3380-473X}, M.~Matthewman, I.~Mikulec\cmsorcid{0000-0003-0385-2746}, J.~Schieck\cmsAuthorMark{2}\cmsorcid{0000-0002-1058-8093}, R.~Sch\"{o}fbeck\cmsAuthorMark{2}\cmsorcid{0000-0002-2332-8784}, D.~Schwarz\cmsorcid{0000-0002-3821-7331}, M.~Shooshtari\cmsorcid{0009-0004-8882-4887}, M.~Sonawane\cmsorcid{0000-0003-0510-7010}, W.~Waltenberger\cmsorcid{0000-0002-6215-7228}, C.-E.~Wulz\cmsAuthorMark{2}\cmsorcid{0000-0001-9226-5812}
% ...
\par}
\vskip\cmsinstskip
\dag:~Deceased\\
$^{1}$Also at Yerevan State University, Yerevan, Armenia\\
$^{2}$Also at TU Wien, Vienna, Austria\\
% ...
\end{sloppypar}
\end{document}